\documentstyle[graphics]{mn}
\newcommand{\kms}{\mbox{km s$^{-1}$}}

\newcommand{\etal}{{et al.}~}

\newcommand{\bc}{\begin{center}}
\newcommand{\be}{\begin{equation}}
\newcommand{\ee}{\end{equation}}
\newcommand{\ec}{\end{center}}

\newcommand{\msun}{\mbox{$M_\odot$}}
\newcommand{\vcirc}{\mbox{$v_{\rm circ}$}}
\newcommand{\Mpc}{\mbox{\rm Mpc}}
\newcommand{\kpc}{\mbox{\rm kpc}}

\newcommand{\vpfit}{{\sc {VPFIT}}}

\newcommand{\lya}{Ly$\alpha$}
\newcommand{\lyb}{Ly$\beta$}

\newcommand{\mnras}{{MNRAS}}
\newcommand{\apjl}{{ApJL}}

\newcommand{\aap}{{A~\&A}}
\newcommand{\NHI}{{\mbox{$N_{\rm H{\sc i}}$}}}

\newcommand{\CIV}{{\mbox{${\rm C}{\sc iv}$}}}
\newcommand{\OVI}{{\mbox{${\rm O}{\sc vi}$}}}

\newcommand{\ltsima}{\mbox{$\; \buildrel < \over \sim \;$}}
\def \simlt{\lower.5ex\hbox{\ltsima}}            % < over ~
\def \gtsima{\mbox{$\; \buildrel > \over \sim \;$}}
\def \simgt{\lower.5ex\hbox{\gtsima}}            % > over ~

\title[Signatures of feedback in QSO absorption spectra] 
{Observational signatures of feedback in QSO absorption spectra}
\author[T. Theuns, H. Mo \& J Schaye]{Tom Theuns$^{1,2}$, H.J. Mo$^2$
and Joop Schaye$^1$\\
(1) Institute of Astronomy, Madingley Rd., Cambridge CB3 0HA, UK\\
(2) Max-Planck Institut f\"ur Astrophysik, Postfach 123, 85740 Garching, Germany}

\begin{document}
\maketitle

\begin{abstract}
Models for the formation of galaxies and clusters of galaxies require
strong feedback in order to explain the observed properties of these
systems. We investigate whether such feedback has observational
consequences for the intergalactic medium, as probed in absorption
towards background quasars. A typical quasar sight-line intersects one
proto-cluster per unit redshift, and significant feedback from forming
galaxies or AGN, heating the proto-cluster gas, will result in a large
clearing of reduced absorption in the \lya-forest. Such a gap could be
detected at redshift $\ga 3$ when the mean opacity is high. Feedback
from Lyman-break galaxies in proto-clusters can be probed by the
absorption lines produced in their winds. Strong feedback from galaxies
has a major impact on the number and properties of absorption lines
with column densities $\NHI\sim 10^{16}$ cm$^{-2}$. This feedback can
be probed with multiple sight-lines and by studying the unsaturated
higher-order lines of the Lyman series. Galactic winds from dwarf
galaxies should break-up into clouds, in order not to over produce the
number of absorption lines. These clouds can then coast to large
distances.
\end{abstract}


\section{Introduction}
In cold dark matter (CDM) dominated cosmogonies structures grow by
gravitational amplification of small primordial density perturbations.
High-density regions decouple from the Hubble expansion and form
gravitationally bound systems with a range of masses up to a
characteristic cut-off mass that increases with decreasing redshift
(Press \& Schechter 1974). Early on, pressure forces are unable to
separate the gas from the dark matter. Galaxies form when baryons that
are dragged into the dark matter potential wells are shocked to
sufficiently high temperatures and densities that the gas can cool
radiatively (Rees \& Ostriker 1977, White \& Rees 1978).

This hierarchical picture of galaxy formation predicts the formation of
too many small systems and therefore conflicts with observations of the
faint-end slope of the galaxy luminosity function. In order to resolve
this discrepancy, feedback is assumed to suppress the formation of most
small galaxies (Larson 1974, Rees \& Ostriker 1977, White \& Rees
1978). The required feedback could be energy input by supernovae (SNe)
associated with star formation (e.g. Dekel \& Silk 1986). In more
massive galaxies, feedback is assumed to be able to prevent the
catastrophic loss of angular momentum of forming discs, which causes
discs in simulations to be much smaller than observed discs
(e.g. Navarro \& Steinmetz 1997, Weil, Eke \& Efstathiou 1998).

On even larger scales, the hierarchical picture based solely on gas
shocking and radiative cooling also seems to conflict with observations
of groups and clusters of galaxies. Without some type of feedback,
shocked gas in the halos of galaxy groups will over produce the X-ray
background (Wu, Fabian \& Nulsen 1999b, Pen 1999). Feedback is also
invoked to explain the observed deviation from self-similarity of the
X-ray properties of hot gas in galaxy clusters (Kaiser 1986, 1991).

Therefore, both theoretical and observational arguments suggest that
forming halos are not just passively accreting material from the
intergalactic medium (IGM), but may be a significant source of energy
input into their surroundings. Is this energy injection a fairly local
process, only affecting gas within a small fraction of the virial
radius of the halo say, or do galaxies stir-up a significant fraction
of the IGM?

This question can be approached from a different angle, by studying the
properties of the IGM in absorption towards background quasars.
Neutral hydrogen along the line of sight produces the \lya~ absorption
lines observed as the quasar's \lq \lya\rq~forest (Bahcall \& Salpeter
1965; Gunn \& Peterson 1965, see Rauch 1998 for a review). Several
theoretical models examined whether collapsed objects could produce the
observed absorption until it was realized that the IGM itself gave rise
to the majority of the lines (McGill 1990, Bi, Boerner \& Chu
1991). Numerical simulations of CDM dominated cosmologies in which the
IGM is photo-ionized by the UV-background from quasars, have been very
successful in reproducing many observed properties of the \lya-forest
(Cen et al. 1994, Zhang, Anninos \& Norman 1995, Hernquist et al 1996,
Theuns et al 1998b; see e.g. Efstathiou, Schaye \& Theuns 2000 for a
review). For example, Theuns et al (1998a) showed that this model
naturally reproduces the evolution in the line densities per unit
redshift and unit column density, from redshift 0 to 4. The superb
quality of the data that this comparison is based on increases our
confidence that the current paradigm of the \lya-forest -- absorption
by neutral hydrogen in the intervening IGM -- is basically correct.

In these simulations of the \lya-forest that do not include any
feedback from forming halos (other than the presence of a
UV-background), strong \lya~absorption lines (column densities $\ge
10^{15}$ cm$^{-2}$) are invariably produced close to forming halos. If
feedback were indeed important, as expected from the earlier
theoretical arguments and from the actual observations of strong
galactic winds associated with star bursts (e.g. Heckman, Armus \&
Miley 1990), one would expect the properties of these stronger lines to
be at least partly influenced by the feedback process.

If galaxy feedback is required to explain the observed X-ray properties
of groups and clusters, then real proto-clusters might be very
different from the proto-clusters in the simulations that do not
include feedback. Consequently, one might expect the absorption
properties of the IGM where the line of sight intersects a proto-cluster
to be rather different in the presence of feedback.

Metal lines (e.g.\ \CIV) associated with $\NHI\sim 10^{14.5}$~cm$^{-2}$
column density lines have been detected in high signal-to-noise QSO
spectra (e.g.\ Cowie et al.\ 1995). Even if the lower-density gas were
polluted by metals at the same level, the associated \CIV\ metal lines
would generally be too weak to be detected. Ellison et al.\ (2000)
searched for \CIV\ in a very high quality spectrum on a pixel by pixel
basis and were able to show that there must be many more metal lines
than can be detected directly. Schaye et al.\ (2000a) performed a
similar search for \OVI\ and found evidence for oxygen enrichment down
to $\tau_{\rm Ly\alpha} \sim 10^{-1}$ at $z \sim 2$--3, which
corresponds to underdense gas.  This is again circumstantial evidence
that a galaxy can influence the gas around it out to a significant
distance. The case is not so clear cut, because the metal pollution
could be the result of an earlier epoch of population III
star-formation. But if galactic winds were the source of the pollution,
then these winds might change the properties of the low density IGM as
observed in absorption.

The aim of this paper is to investigate to what extent the current
picture of the \lya-forest can incorporate a significant amount of
feedback, without destroying the good agreement with observations. In
what follows we will also use numerical simulations of the
\lya-forest. We simulate a vacuum energy dominated, flat cold dark
matter model, with density parameters
$(\Omega_m,\Omega_\Lambda)=(0.3,0.7)$ (e.g. Efstathiou 1999 and
references therein), Hubble constant $h=0.65$ (Freedman et al. 1999)
and baryon density $\Omega_b h^2=0.019$ (Burles \& Tytler 1998) of
which a fraction $Y=0.24$ by mass is helium, normalized to the
abundance of clusters (Eke, Cole \& Frenk 1996). We assume the IGM to
be photo-ionized and photo-heated and chose the spectrum of the
ionizing sources such that the simulation reproduces the thermal
evolution and mean absorption as determined from observations by Schaye
et al. (2000b; the gas temperature as function of redshift follows the
dashed line in Figure~6 of that paper). The cubic simulation box is
20~Mpc on a side (co-moving) and contains $256^3$ particles of each
species (i.e. dark matter and gas particles; note that the designer
model in Schaye et al.  (2000b) was a smaller simulation but with the
same thermal history), providing sufficient resolution to simulate the
gas reliably (Theuns et al. 1998b). Without feedback, this simulation
reproduces most of the observed properties of the \lya-forest, for this
choice of currently popular values of the cosmological parameters. it
includes star-formation, whereby gas at a density contrast
$\rho>80\langle\rho\rangle$ is converted to \lq stars\rq~, when the
temperature drops below $2\times 10^4$K. No feedback (other then a
UV-background) is included.

This paper consists of two parts. In the first part we assume that the
gas in a proto-cluster is pre-heated at some early time, and discuss
the corresponding observational absorption signature if the line of
sight to a quasar intersects the proto-cluster. In the second part, we
consider feedback effects from smaller systems and examine their impact
on the IGM.

\section{Feedback on cluster scales}
We begin this section by briefly reviewing the arguments that suggest
that the entropy of proto-cluster gas was increased significantly by
some non-gravitational process at an early time $z_{\rm inj}\ge 2$. We
then argue that this gas is too hot to be confined by a galactic
potential well and hence will fill a considerably fraction of the
volume of the proto-cluster. The probability that a quasar sight-line
intersects such a hot bubble is $\sim 1$ per unit redshift, and it will
appear as a region of decreased absorption in the \lya-forest. This
should be relatively prominent especially at high redshift $z\ge 4$
when the mean \lya-opacity is high. Other processes, such as proximity
effects from AGN and/or galaxies in the proto-cluster, could contribute
to clearing a gap in the absorption. Observations of metal absorption
lines can be used to examine the thermal properties of the medium, and
hence distinguish a proto-cluster gap from a void, and make inferences
about the nature of the sources producing the feedback. If the
proto-cluster gas is in a two-phase medium, with cold clouds
pressure-confined in the hot phase, then the gaps may not be completely
empty.

\subsection {General considerations}
Clusters of galaxies are the most massive virialised systems in the
universe, with typical total masses of about $10^{15}\msun$, of which
10--30 per cent is baryonic. The latter component is predominantly in
the form of a hot X-ray emitting intra-cluster medium (ICM), with
temperature and X-ray luminosities typically of order $T\sim 10^8{\,\rm
K}$ and $L_X\sim 10^{45}\,{\rm erg\,s^{-1}}$, respectively.

The intracluster gas is believed to be shock-heated as it falls into
the cluster potential well. If the initial gas temperature is
negligible, then a relation $L_{\rm X}\propto T^2$ is predicted from
simple scaling arguments (Kaiser 1986), which disagrees with the
observed result, $L_{\rm X}\propto T^\alpha$, with $\alpha=2.6$--3
(Jones \& Forman 1984, David, Forman \& Jones, 1991, Lloyd-Davies,
Ponman \& Cannon 2000). Consequently, it was suggested that the ICM gas
is \lq pre-heated\rq~ by some non-gravitational process (Kaiser 1991;
Evrard \& Henry 1991; Navarro, Frenk \& White 1995; Cavaliere, Menci \&
Tozzi 1997, Ponman, Cannon \& Navarro 1999). The existence of such an
entropy \lq floor\rq~ breaks the self-similarity and helps to explain
the $L_{\rm X}$--$T$ relation (Kaiser 1991). Since an initial entropy
is expected to have a larger effect in clusters with lower temperature,
the level of pre-heating may be revealed by measuring the specific
entropy of gas in X-ray clusters as a function of cluster
temperature. The effect of pre-heating would be largest in groups of
galaxies, which would emit far less X-rays when pre-heated since the
hotter gas is less dense, thereby avoiding the problem of over
producing the X-ray background (Wu, Fabian \& Nulsen 1999b, Pen 1999).

Defining a specific \lq entropy\rq
\begin{equation}
\label{s_defined} s\equiv {T\over n_{\rm e}^{2/3}}\,,
\end{equation}
where $T$ is the temperature of the gas and $n_{\rm e}$ is the electron
number density, then the level of pre-heating inferred from the
observations is on the order of
\begin{equation}
\label{s_observed}
s_0\sim 100 h^{-1/3} {\,\rm keV\, cm^2}\,
\end{equation}
(Ponman, Cannon \& Navarro 1999, Lloyd-Davies, Ponman \& Cannon 2000,
Balogh, Babul \& Patton 1999, Wu, Fabian \& Nulsen 2000). We will use
$s_{100}\equiv s_0/(100 h^{-1/3}{\,\rm keV\,cm^2})$.

At present it is not yet clear which sources dominate the heating of
the ICM. The {\em energy} required to reach a given entropy floor
increases with the gas density as $E\propto n_e^{2/3}$, which makes it
easier energetically to heat the gas before it collapses into a
potential well. If supernova explosions and stellar winds,
e.g. associated with episodes of rapid star-formation, are the main
sources of energy, then pre-heating is likely to have occured
relatively early on. Indeed, recent observations show that large
numbers of stars are being formed at $z\ga 3$ (Steidel et al. 1996) in
strongly clustered galaxies (e.g. Giavalisco et al. 1998, Adelberger et
al.  1998) which are likely progenitors of the bright galaxies in
present-day clusters (Mo \& Fukugita 1996, Baugh et al. 1998, Wechsler
et al. 1998, Mo, Mao \& White 1999).  Semi-analytical models for the
formation of cluster elliptical galaxies also predict that most of
their stars formed at $z\ga 2$ (Kauffmann, White \& Guiderdoni 1993,
Kauffmann et. al. 1999). In addition, Mushotzky \& Loewenstein (1997)
argued that the observed lack of evolution in the iron abundance of
cluster gas out to $z\sim 1$ also suggests that enrichment occured
early. If, on the other hand, energy input from AGN is the main energy
source, then pre-heating is also likely to have occured at $z\ga 2$,
because the number density of QSOs peaks around $z=2$-3 (Shaver et
al. 1996). Therefore, for both types of potential energy sources, stars
or AGN, we expect that the major part of pre-heating took place before
$z\sim 2$.

The inferred entropy floor Eq.~(\ref{s_observed}) implies that the ICM
was heated to some high temperature at an early time. At a redshift
$z$, the number density of electrons, $n_e$, of fully ionized
proto-cluster gas of primordial abundance (with a helium fraction of 24
per cent by mass), at over density $\delta$ is
\begin{equation}
n_e\approx 1.2\times 10^{-5} (1+\delta)
\left({\Omega_b h^2\over 0.019}\right)\left({1+z\over 4}\right)^3\,.
\end{equation}
In order to achieve the required entropy~(\ref{s_observed}), the gas in
the region must be heated to a temperature
\begin{eqnarray}
\label{T_required}
T\!\!&\approx &\!\!0.06 \left({s_{100}\over f_s^{1/3}
\phi^{2/3}}\right) (1+\delta)^{2/3} \left({\Omega_b h^2\over
0.019}\right)^{2/3}\nonumber\\ &\times&\left({1+z\over
4}\right)^2\left({h\over 0.65}\right)^{-1/3}\,{\rm keV}\,,
\end{eqnarray}
where $f_s$ and $\phi$ are the heated mass and volume fractions of gas,
respectively. The implied temperature is quite high ($T\ga 10^6$\,K;
1keV$\approx 1.2\times 10^7$K) if pre-heating occured at $z>2$, even if
the heating occured close to the mean density $\delta\sim 0$. At $z<1$,
the required $T$ is lower ($T\sim 10^5$\,K) if $\delta=0$.  However,
for a cluster to form at the present time, the redshift at which it
turned around is about 1, in which case the gas density when the
pre-heating took place must have been $\delta\sim 10$ over the volume
of the proto-cluster, so we again obtain $T\ga 10^6$\,K.

If pre-heating occured inside galaxy halos, then $f_s$ and $\phi$ are
small and the required temperature Eq.~(\ref{T_required}) is higher
than the virial temperature of the halos. Such hot gas cannot be
confined to a galaxy halo and therefore, irrespective of the heat
source, the pre-heated gas must have quite an extended distribution,
filling a large fraction of the region occupied by a proto-cluster.

If pre-heating is due to SNe, one can estimate the total amount of stars
needed to heat the proto-cluster to the required temperature. Suppose
the gas mass of the proto-cluster is $M_{\rm gas}$, and $f_\star M_{\rm
gas}$ is the mass of stars responsible for pre-heating. The total energy
released by the SNe associated with those stars is
\begin{equation}
E={M_{\rm gas}\over \nu} f_\star E_{51} 10^{51}\,{\rm erg}\,,
\end{equation}
where $E_{51}$ is the energy output per SN in units of $10^{51}\,{\rm
erg}$ and $\nu$ is the stellar mass needed to produce one supernova (we
will write $\nu=100 \nu_{100}\msun$). If a fraction $\epsilon$ of
all of the SN energy is available to heat the gas, then the gas
temperature will be
\begin{equation}
kT \approx 2.1 \epsilon \left({E_{51} f_\star\over \nu_{100}}\right)
\,{\rm keV}\,.
\end{equation}
It then follows that $f_\star$ must be about $0.1 \epsilon^{-1}$ to
$0.2\epsilon^{-1}$ in order to heat the proto-cluster gas to a few
million degrees. This fraction is consistent with the observed fraction
of baryonic mass in stars ($\sim 10$ per cent) and the metallicity of
X-ray emitting gas in clusters, if the efficiency is high,
$\epsilon\sim 1$.  However, if only a small part of the feedback energy
goes into heating the gas, then supernova explosions and stellar winds
may not be sufficient to heat the gas to the required temperature, and
heating from AGN may have to be invoked (e.g. Wu, Fabian \& Nulsen
2000).

Another requirement for pre-heating to work is that the heated gas does
not lose the injected entropy radiatively. The time scale $t_{\rm
cool}\equiv T/\dot T$ for radiative cooling is
\begin{eqnarray}
\label{coolingtime}
t_{\rm cool} &\approx& 1.0\times 10^{11} \sqrt{T_6}\Lambda_{-23}^{-1}
\left({\Omega_b h^2\over 0.019}\right)^{-1} (1+\delta)^{-1}\nonumber\\
&\times&\left({1+z\over 4}\right)^{-3}\, {\rm yr}\,,
\end{eqnarray}
where $T_6$ is the gas temperature in $10^6$\,K and $\Lambda(T)\equiv
\Lambda_{-23}\times 10^{-23}\,\sqrt{T_6}{\rm\,erg\,cm^3\, s^{-1}}$ is
the radiative cooling rate.  For T$\ge 10^6$K, cooling is dominated by
thermal Bremsstrahlung and $\Lambda_{-23}\sim 1$, whereas at lower
temperatures $\ge 10^4$K line-cooling becomes important and the rate
can increase up to $\Lambda_{-23}(T)\sim 10^2$ depending on the
metallicity and the temperature of the gas. At high temperatures and
for tenuous gas, cooling will be dominated by Compton losses on the
micro-wave background radiation, and the cooling time $t_{\rm cool}\sim
1\times 10^{10}\,[(1+z)/4]^{-4}$ yr. Comparing these times with the
Hubble time
\begin{equation}
t_{\rm H}\approx 3.4\times 10^{9}(h/0.65)^{-1}
\left(\Omega_m/0.3\right)^{-1/2} \left({1+z\over 4}\right)^{-3/2}{\rm
yr}\,,
\label{eq:thub}
\end{equation}
(here and in what follows, we use the high-redshift approximation
$H(z)\approx H_0 \Omega_m^{1/2} (1+z)^{3/2}$, valid for a
cosmologically flat universe at $z\ge 2$) and using
Eq.~(\ref{T_required}), we see that the cooling time is equal to the
age of the Universe when $\delta\sim 100$ at $z=3$. Again this shows
that pre-heating should occur in gas at low over density, since
otherwise the injected entropy will be lost radiatively.

\subsection {Observational signatures}
\begin{figure*}
\setlength{\unitlength}{1cm}
\centering
\begin{picture}(21,6)
\put(0., -12){\includegraphics{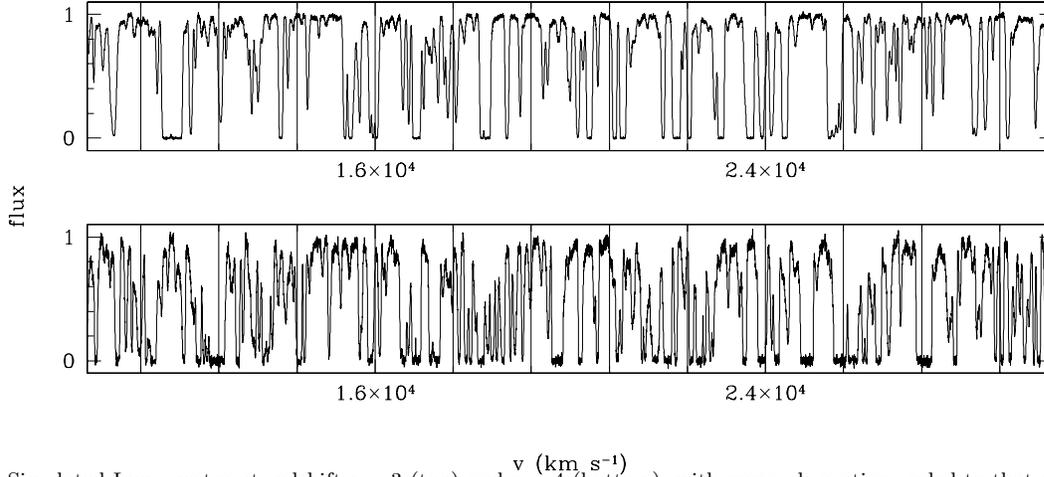}}
\end{picture}
\caption{Simulated \lya~spectra at redshift $z=3$ (top) and $z=4$
(bottom), with mean absorption scaled to that of observed spectra
(Schaye et al 2000b). Vertical lines are drawn at intervals given by the
expected size of a cluster, Eq.~(\ref{eq:dV}). Intervals with little
absorption over a region the size of a proto-cluster are very rare,
especially at $z=4$ when the mean opacity is high. A spectrum of this
length contains of order one proto-cluster, which would stand-out as a
region of decreased absorption if the proto-cluster gas were hot due to
entropy injection.}
\label{fig:cluster}
\end{figure*}
In the previous section we argued that pre-heating of gas in clusters
is likely to have occured before $z\ge 2$, and that the gas temperature
is fairly high, $T\ga 10^6$K.  We also argued that the hot gas should
fill a relatively large fraction of the volume occupied by the
proto-cluster, since the hot gas in unlikely to be confined by a
galactic potential well. Because of the high temperature, the gas in a
proto-cluster will be collisionally ionized thereby producing a region
of reduced absorption in a QSO sight-line that intersects the
cluster. In addition, proximity effects due to AGN and/or galaxies may
further reduce the amount of absorption. What are the expected sizes
and rates of incidence of such \lq gaps\rq?

The rate of incidence of objects with co-moving density $n$ and
physical radius $R$ is
\begin{equation}
{dN\over dz} = n\pi R^2 {(1+z)^2\over H(z)}\,.
\label{eq:dNdz}
\end{equation}
The physical radius of a proto-cluster of mass $M=10^{15} M_{15}h^{-1}\msun$,
which is at an over-density $\delta$ at redshift $z$, is
\begin{equation}
R\approx 10 \Omega_m^{-1/3}(1+\delta)^{-1/3} (1+z)^{-1} M_{15}^{1/3}
h^{-1}{\,\rm Mpc}\,,
\label{eq:rclus}
\end{equation}
and consequently the average number of times a sight-line will
intersect such a a proto-cluster is (for $z\ga 2$)
\begin{equation}
\label{eq:dNdz_defined}
{dN\over dz} \approx 1.9\, n_4 (\Omega_m/0.3)^{-7/6} (1+\delta)^{-2/3}
M_{15}^{2/3} \left(1+z\over 4\right)^{-3/2}\!\!\!\!\,,
\end{equation}
where $n_4$ is the co-moving number density of proto-clusters in units
$4\times 10^{-6} \,h^3{\rm Mpc}^{-3}$. For Abell clusters with mass
$M_{15}\ga 0.5$, $n_4\sim 1$.

The differential Hubble velocity $V=H(z)R$ over a region of this size
is
\begin{eqnarray}
V &=& 1.6\times 10^3\,\left(\Omega_m/0.3\right)^{1/6}
\left(1+z\over 4\right)^{1/2}\nonumber\\
&\times & (1+\delta)^{-1/3} M_{15}^{1/3}
\kms\,,
\label{eq:dV}
\end{eqnarray}
and the corresponding wavelength range in a \lya~spectrum is
\begin{equation}
\Delta\lambda\approx 26 \left(\Omega_m/0.3\right)^{1/6}\left(1+z\over
4\right)^{3/2} (1+\delta)^{-1/3} M_{15}^{1/3}{\rm \AA}\,.
\label{Deltalambda}
\end{equation}
The observational signature of a proto-cluster will be smaller than
this, since peculiar velocities will compress the cluster's extent in
redshift space. The size of the cluster given by
Eq.~(\ref{Deltalambda}) is large compared to the distance between lines
at $z\sim 3$, and even more so at higher redshifts where the mean
opacity in the forest is higher, as demonstrated in
Fig.~\ref{fig:cluster}. This figure shows a simulated \lya-spectrum of
a $z_{\rm em}=3$ and a $z_{\rm em}=4$ quasar, extending over a velocity
range typical for observed quasars. (Here and below, velocity $v$ is
related to wavelength $\lambda$ by $dv/c=d\lambda/\lambda$). Such
spectra do contain \lq gaps\rq~ with little absorption, which are due
to large voids resulting from long-wavelength perturbations, but only
very rarely are such voids of the size given by Eq.~(\ref{eq:dV}),
especially at the higher redshift where the mean absorption is
higher. The simulation used to make Fig.~\ref{fig:cluster} lacks
large-scale modes due to it's finite size -- 20~Mpc -- and this may
lead to an underestimate of the sizes of the largest voids. A more
quantitative comparison against data could take such long-wavelength
modes into account by using for example the log-normal model of Bi \&
Davidsen (1992).

Thus, if pre-heating of the ICM occurred at $z\ga 3$, its signature may
be apparent in a QSO spectrum that intersects a
proto-cluster. Therefore, QSO absorption line spectra are potentially
powerful probes of feedback on cluster scales, since an average quasar
sight-line will intersect a proto-cluster region of order $\sim 1$
times, according to Eq.~(\ref{eq:dNdz_defined}).

However, there are complications in the model prediction. At low
redshifts $z\le 1$, proto-clusters may have already began to contract
relative to the Hubble flow, and the clearings in the forest produced
by the hot regions in QSO spectra are smaller than when assuming pure
Hubble flow.  Since the line density in the \lya-forest decreases with
redshift, it becomes increasingly difficult to associate a clearing in
the forest with a proto-cluster, since such gaps naturally appear as a
consequence of the emptying of voids and the overall decrease in the
\lya~optical depth.

Another complication is that some cold gas may exist in the hot
proto-clusters and produce \lya~absorption lines, so the clearings may
not be empty. Cold gas\footnote{The sight-line may also intersect a
galaxy. If the cluster contains $10^3$ galaxies, then the covering
factor of intersecting one within $30h^{-1}\kpc$ is $10^3\times
[0.03(1+z)/10]^2\sim 0.1$ for a $R=10h^{-1}$Mpc cluster at $z=3$.} is
likely to co-exist with the hot phase in the form of pressure-confined
clouds, in analogy with the multi-phase model of the interstellar
medium (ISM, McKee \& Ostriker 1977). In these models of the ISM, cold
clouds are pressure-confined by a hot medium generated by supernova
explosions. Such models have also been considered for the \lya-forest
itself (Ostriker \& Ikeuchi 1983) as well as for the QSO absorption
systems associated with galaxies (Mo 1994; Mo \& Miralda-Escud\'e
1996).

We can estimate the properties of these cold clouds, by assuming
pressure equilibrium between the hot and cold phases,
\begin{equation} 
n_{\rm c} T_{\rm c}=n_{\rm h} T_{\rm h}\,. 
\end{equation}
If the cold clouds are heated by the general UV-background, their
temperatures will be typically $\sim 10^4{\,\rm K}$. If $f_{\rm
c}\equiv M_{\rm c}/M_{\rm gas}$ is the fraction of gas in the cold
phase, and $m_{\rm c}$ is the typical mass of a cold cloud, then
\begin{equation}
C=f_{\rm c}\left[\left({n_{\rm h}\over n_{\rm
c}}\right)^2 {M_{\rm gas}\over m_{\rm c}}\right]^{1/3}\,
\label{eq:cover}
\end{equation}
is the covering factor of the clouds.  The mass of cold clouds in a hot
medium is limited by various physical processes (evaporation,
hydrodynamical and gravitational instabilities; see e.g. Mo \&
Miralda-Escud\'e 1996), and typical values are $m_{\rm c}\sim
10^6\msun$. Equation~(\ref{eq:cover}) shows the covering factor to be
$\sim 1$ requires that about 10 per cent of the proto-cluster gas is in
the cold phase. In that case, each time we observe a clearing in the
forest, produced by a proto-cluster, one absorption line produced by
the pressure-confined clouds is expected. If this component of
absorption lines is indeed important, the proto-cluster gap in the
spectrum would be smaller.

The HI column density of a pressure-confined cloud is in the optically
thin case
\begin{eqnarray}
N_{\rm HI}&\approx& 1.6\times 10^{15} {R\over 1\kpc}
\left({P\over 20{\rm
K\,cm^{-3}}}\right)^2\nonumber\\
&\times& \left({T_c\over 10^4
{\rm K}}\right)^{-11/4} J_{21}^{-1} {\rm cm}^{-2}\,,
\end{eqnarray}
where $P$ is the pressure of the confining medium, $R$ is the cloud's
radius, $T_c$ is the temperature of the cold medium, and $J_{21}$ is
the ionization flux at the Lyman limit, in the usual units of
$10^{-21}\,{\rm erg\,s^{-1}\,cm^{-2}\,sr^{-1}\,Hz^{-1}}$ (see Mo \&
Morris 1994). For typical values $P\sim 20$ (corresponding to $T_{\rm
h}=10^{6}{\,\rm K}$ and $n_{\rm h}\sim 2\times 10^{-5}{\,\rm
cm}^{-3}$), the size of a cold cloud (with mass $10^6\msun$) is about 1
kpc, and the typical HI column density of such cloud is thus $N_{\rm
HI}\sim 10^{15.5}{\,\rm cm}^{-2}$. Therefore these pressure-confined
clouds can produce strong \lya~absorption lines. Since small and
diffuse clouds will evaporate in the hot environment, weak \lya~lines
are not expected in the hot gap.

We will show below (Fig.~\ref{fig:nh16}) that the typical strong lines
of column density $N_{\rm HI}\sim 10^{16}{\,\rm cm}^{-2}$ produced in
simulations of the \lya-forest are the result of the overlap in
redshift space of several filaments. The \lya-profile of such a line is
saturated but the substructure can still be probed by the higher-order
lines (e.g. \lyb). In addition, the extent of such absorbers {\em
perpendicular} to the line of sight, which can be probed using double
or lensed quasars, is very much larger than for the $\sim 1$\kpc~
pressure-confined clouds. Therefore it is possible to distinguish
strong lines produced by a single cold cloud, as described above, from
those that are typical in gravitational structure formation.

If pre-heating is of stellar origin, then the ICM may be enriched in
metals by supernova explosions and stellar winds. The mass-weighted
mean metallicity of the proto-cluster gas is $Z\sim y f_\star$, where
$y$ is the metal yield per unit stellar mass. Since $y\sim Z_\odot$, we
have $Z\sim f_\star Z_\odot$, i.e. the proto-cluster medium can be
enriched to a level of $0.1$--$0.2\, Z_{\odot}$, consistent with the
observed metallicity of the present ICM (e.g. Renzini 1997). The
absorbing clouds confined in the ICM are presumably enriched to a
similar level. If this is the case, we would expect to observe strong
metal lines associated with the \lya~clouds.  Using the typical density
and size of the cold clouds, and assuming the UV-background intensity
has a power law form $J(\nu)\propto \nu^{-1}$ with amplitude given by
$J_{21}\sim 1$, the associated CIV absorption systems are expected to
have column densities of $10^{13}$--$10^{14}{\,\rm cm}^{-2}$,
comparable to the observed CIV column densities associated with the
strong lines in the \lya-forest (e.g. Cowie et al. 1995). If the
hot ICM is polluted by metals as well, then a proto-cluster gap may be
associated with very broad metal lines, due to the high gas temperature
and differential Hubble expansion across the region. Such a
characteristic signature may help in distinguishing a proto-cluster gap
from a void.

If, on the other hand, heating and metal enrichment have different
origins (e.g., heating was done by AGN), metals produced by supernovae
and stellar winds may be confined to small regions around individual
galaxies. In this case, a large fraction of the heated ICM may remain
at low metallicity, and the strong \lya~lines produced by
pressure-confined clouds may not be associated with any strong metal
lines. Thus, by measuring the metallicities of the strong lines
produced in the proto-cluster regions, we may hope to obtain
constraints on the sources responsible for the pre-heating.
Proto-cluster regions may be identified as large concentrations of
galaxies in the galaxy distribution.  Analyses of QSO absorption
spectra behind such regions are therefore important.

Since the ICM is confined in the proto-cluster potential, the gas will
move with the dark matter as the proto-cluster expands and collapses. We
therefore do not expect to see wind-driven shells to form on the
boundary of a proto-cluster.

\subsection {Connection with Lyman-Break galaxies}
\label{sect:lbg}
Large numbers of star forming galaxies have been discovered at $z\ga 3$
by colour selection using the Lyman-break technique (Steidel et
al. 1996).  The star-formation rates in these galaxies, as inferred
from their UV-luminosities, are in the range 10 to $1000\msun{\,\rm
yr}^{-1}$ (for a flat universe with $\Omega_m=0.3$ and for a mean dust
correction of $\sim 7$). The strong clustering of these galaxies is
consistent with many of them being the progenitors of the early-type
galaxies as seen in clusters today (e.g. Mo \& Fukugita 1996; Governato
et al 1998). Thus, if pre-heating of the proto-cluster medium occured
at $z\ga 3$, we may hope to see the kind of signatures discussed in the
last subsection in the spectra of QSOs which lie behind a large cluster
of LBGs.
 
There is an interesting, although indirect, argument (Pagel 1999; see
also Pettini 1999) that the metals returned from LBGs are mostly in hot
gas.  Indeed, if all the metals ejected from LBGs were in cold gas,
they should be observed as metal line systems associated with \lya~
absorption systems. However, the observed total amount of metals in
\lya~absorption systems is much lower than that expected from the star
formation inferred to occur in LBGs.  A possible solution is that most
metals are hidden in hot gas which would not produce significant \lya~
absorption. If this is the case, then metals are most likely contained
in the proto-cluster medium, rather than being confined around
individual galaxies, because radiative cooling would be very efficient
if the gas were confined to such high density regions [see equation
(\ref{coolingtime})].  Alternatively, the apparent scarcity of metals
could be a selection effect, when metal-rich LBGs block background
quasars due to dust obscuration. Clearly, observations of the
proto-cluster medium around LBGs can help to distinguish between these
two possibilities.

If metal-enriched gas is being ejected from LBGs into the proto-cluster
medium by winds, then we may hope to observe the signature of such gas
outflows. Because of their high star-formation rate, LBGs are
considered to be the high-redshift counterpart of local star burst
galaxies (e.g. Heckman 2000).  Therefore, we can use observed
properties of local star bursts to predict the observable signatures of
LBGs in the QSO absorption line spectra. As summarized in Heckman et
al. (1999), local star bursts are regions of intense star-formation
(more than $10\msun$ stars are being formed per year, in a region of
$\sim 1$ kpc), which generally show gas outflows with velocities in the
range $400$--$800\kms$. These high outflow velocities do not strongly
depend on the depth of the potential well (Heckman 2000), and are
sufficiently high that a galaxy can in principle pollute a large volume
of the IGM in a Hubble time. The mass outflow rate is roughly
proportional to the star-formation rate in the galaxy, ${\dot M}_{\rm
w}\propto {\dot M}_{\star}$, with the proportionality factor being
typically a few. The wind consists mainly of cold clouds entrained in
the outflowing hot gas. The ratio between the mass outflow rate and the
rate at which mass is being returned by supernova explosions and
stellar winds is typically 5 to 10. However, only about 10 percent of
the energy output from supernova explosions and stellar winds is in the
kinetic energy of the cold flow; the rest is in the thermal and kinetic
energy of the hot wind.

We can construct a simple model (see also Wang 1995ab) for the
properties of the winds around LBGs, assuming them to have similar
outflow properties as star bursters. The density structure in a steady
wind is
\begin{equation}
\rho_{\rm w}(r)={{\dot M}_{\rm w}\over \omega r^2 v_{\rm w}}\,,
\end{equation}
where $v_{\rm w}$ is the wind velocity and $\omega$ is the solid angle
into which the wind blows. Inserting numerical values typical for a
star burst wind gives for the hydrogen number density:
\begin{eqnarray}
n_{\rm H}(r)\!\!\!\!&\sim& 1\times 10^{-2}\left({\omega\over 4\pi}\right)^{-1}
\left({{\dot M}_{\rm w}\over 100\msun\,{\rm yr}^{-1}}\right)
\left({v_{\rm w}\over 200\kms}\right)^{-1} \nonumber\\
&\times &\left({r\over
10\kpc}\right)^{-2}{\rm cm}^{-3}\,.
\end{eqnarray}
In the presence of an ionizing background, the neutral fraction of gas
with clumping factor $\zeta\equiv\langle n_{\rm HII}^2\rangle/ \langle
n_{\rm HII}\rangle^2 $ is
\begin{equation} 
{n_{\rm H{\sc I}} \over n_{\rm H}} \sim 4\times 10^{-4} T_4^{-3/4}
{\zeta\over J_{21}} \left({n_{\rm H}\over 10^{-3}\,{\rm
cm}^{-3}}\right)\,,
\label{eq:peq}
\end{equation}
where $T_4\sim 1$ is the temperature of the photo-ionized gas in units
of $10^4$\,K, and the spectrum of the ionizing radiation is $\propto
\nu^{-1}$ (e.g. Peebles 1993, section 23).  Combining these two
equations, we find for the neutral hydrogen column density at an impact
parameter $l$
\begin{eqnarray} 
N_{\rm H{\sc I}}&\sim& 1\times 10^{15} 
T_4^{-3/4} {\zeta\over J_{21}}
\left({\omega\over 4\pi}\right)^{-2}\left({{\dot M}_{\rm w}\over 100\msun\,{\rm
yr}^{-1}}\right)^2\nonumber\\
&\times &\left({v_{\rm w}\over 200\kms}\right)^{-2}\left({l\over 50\kpc}\right)^{-3}
{\rm cm}^{-2}\,.
\label{eq:nwind}
\end{eqnarray}
Thus, a QSO sight-line passing near a LBG will show a relatively
strong, possibly broad \lya~line, associated with the wind driven by
the star burst. Complex velocity structure is to be expected if the
outflow is clumpy. For an impact parameter smaller than $\sim 10\kpc$,
the QSO sight-line goes through optically thick clouds, and a Lyman
limit system or a damped \lya~system will be produced.

The observational results by Pettini et al. (2000) for the spectrum of
the gravitationally lensed galaxy MS1512-cB58 (at $z=2.73$) give direct
evidence for such gas outflows with velocities $\sim 200\kms$ in
LBGs. However, common LBGs are too faint to yield spectra with high
signal-to-noise ratios. Such outflows may be studied more efficiently
is absorption towards background QSOs.

Because of their large velocities, winds from LBGs may drive shocks
into the proto-cluster medium as they escape the potential wells of
their parent halos. In the adiabatic phase, the shock front has a
density contrast of about 4 with respect to the ambient medium and a
temperature $T\sim \mu v_{\rm w}^2/2k\sim 2\times 10^6 (v_{\rm
w}/200\kms)^2\,{\rm K}$, where $\mu$ is the mean molecular weight of
the gas. Using equation (\ref{coolingtime}) we see that the shocked gas
will remain very hot, as it is not be able to cool within a Hubble
time. Thus, the winds may drive hot shells into the medium, which will
eventually thermalize in the ambient proto-cluster gas. Since the gas
in these shells is collisionally ionized, they are not expected to
produce strong lines.

In summary: a typical $z\sim 3$ quasar spectrum will intersect one
proto-cluster per unit redshift. If the gas in this cluster is hot, as
is expected from models that invoke pre-heating to explain the observed
X-ray properties of clusters, then a large region with little \lya~
absorption would result, which might also produce associated metal
absorption in the form of Hubble broadened metal lines if the ICM is
metal enriched. Such clearings in the forest would be apparent at
sufficiently high redshifts, when the mean opacity in the \lya-forest
is high. Proximity effects, due to AGN and/or galaxies, might make the
clearing even more distinct. However, if the proto-cluster gas is in a
two-phase medium, with cold clouds pressure-confined in the hot medium,
then these clouds may produce strong lines themselves. These could be
distinguished from the more usual strong lines by the absence of
substructure, probed by higher-order Lyman lines, or by their small
sizes, as measured from multiple quasar sight-lines. If the cold phase
is metal enriched there would be a strong metal line associated with
these lines. If proto-clusters usually contain a large number of
Lyman-break galaxies, then QSO spectra taken through such a
concentration can be used to test these predictions. Winds from LBGs
are also expected to produce strong lines with column densities $>
10^{15}$ cm$^{-2}$ out to large impact parameters. These lines may have
associated metal absorption if the wind is enriched.

\section{Feedback from small systems}
Feedback on the scale of galaxy halos appears to be required for a
variety of reasons. Without feedback, a major fraction of all baryons
will cool and form stars in small halos at high redshift, exhausting
the baryon supply to make up the IGM and the discs at lower
redshift. This is known as the over cooling problem. The presence of an
ionizing background, for example due to AGN, may suppress the collapse
of baryons into dark halos of circular speeds below $\sim 20-30$ \kms
(Efstathiou 1992, Quinn, Katz \& Efstathiou 1996; Thoul \& Weinberg
1996). For more massive systems, feedback associated with
star-formation is usually invoked to explain the low baryon content of
dwarf galaxies and also to prevent forming galactic discs from loosing
most of their angular momentum to their dark halo (Weil, Eke \&
Efstathiou 1998).

The basic picture for star-formation feedback is that hot bubbles blown
by SNe and stellar winds overlap and power a galactic wind. Cosmic rays
produced in the SNe shocks may also contribute (Breitschwerdt, McKenzie
\& V\"olk 1991). This feedback process heats cold gas which is then no
longer available for star-formation, and may blow some fraction of the
baryonic material out of the virial radius of the halo. In the next
step of the merging hierarchy, the blown-out gas may once more be
accreted.

Semi-analytical models of galaxy formation (e.g. Kaufman, White \&
Guiderdoni 1993, Lacey et al. 1993) require that a significant fraction
of the total energy associated with SNe explosions be converted into
thermal energy of the heated gas. However, numerical simulations of
this process (e.g. Navarro \& White 1994) fail to reach the required
efficiency, because SN explosions invariably occur in dense gas which
rapidly cools radiatively. Presently it is not yet clear whether the
high efficiencies assumed in the semi-analytic models are unrealistic,
or whether the simulations grossly underestimate the efficiency because
they lack the resolution to produce a multi-phase medium.

Efstathiou (2000) recently proposed a model where feedback produces a
multi-phase interstellar medium in which a low rate of star-formation
can produce steady conversion of cold gas into hot gas, which then
drives a wind. In this model, star formation is quenched because SNe
explosions limit the amount of cold gas available for star-formation.
The model shows that a galaxy with circular speed $\sim 50$ km s$^{-1}$
can expel 60-80 per cent of its gas over a time-scale of $\sim$ 1
Gyr. Nath \& Trentham (1997) also investigated a model where an early
population of dwarfs underwent strong mass loss through winds, and
showed that this could pollute the low density IGM to a value close to
those inferred from observations. The details of how such a wind
interacts with the infalling gas may be very important for a realistic
description of the feedback process, and also for estimating to what
extent a forming galaxy can influence its surroundings. For example,
Efstathiou (2000) argued that the wind might break-up into clouds which
do not interact strongly with the accreting gas. Therefore, such a
galaxy would undergo inflow and outflow simultaneously.

In the next sections, we use our simulations to investigate two very
simple models for possible effects of galaxy feedback on the
\lya-forest. In the first model we remove all the gas in the
surroundings of galaxies so that it no longer contributes to the
absorption. In this \lq maximal feedback\rq~ model, we assume that
galactic outflows are very powerful and heat-up a significant fraction
of the IGM surrounding them. We will show that this has a dramatic
effect on the number of strong absorption lines. In the next section,
we examine the effects of galactic winds around dwarfs on the IGM. We
show that shells formed from such a wind will over produce the number
of \lya-lines, unless the wind breaks-up into clouds.

\subsection{Feedback and strong lines}
\begin{figure*}
\setlength{\unitlength}{1cm}
\centering
\begin{picture}(21,4)
\put(0., -14){\includegraphics{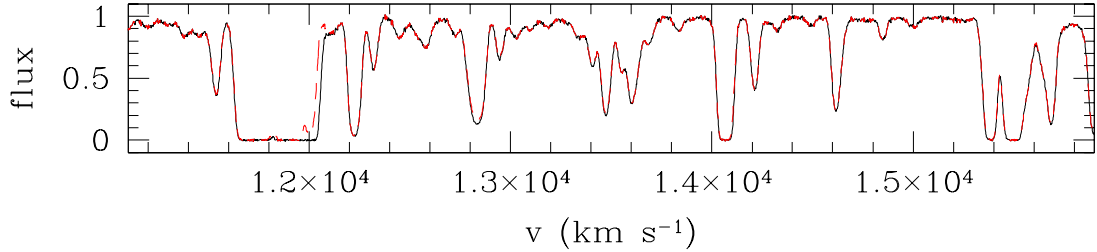}}
\end{picture}
\caption{Simulated \lya~spectrum with (full line) and without (dashed
line) the contribution from hot gas (see text for definition of \lq
hot\rq). This stretch was carefully picked to show any difference at
all between these two cases: hot gas contributes very little to the
observed spectrum, as expected.}
\label{fig:spec2}
\end{figure*}
\begin{figure*}
\setlength{\unitlength}{1cm}
\centering
\begin{picture}(21,8)
\put(0., -10.5){\includegraphics{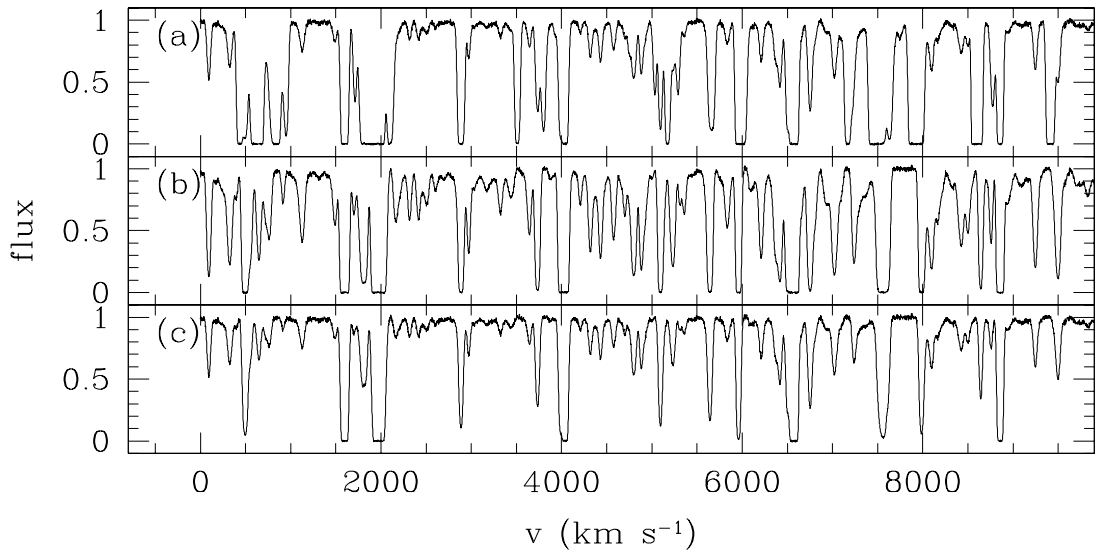}}
\end{picture}
\caption{Simulated \lya~spectra at $z\approx 3$.  Panel (a): simulated
spectrum without feedback; panel (b): same as panel (a) but with
feedback and scaled to have the same mean absorption as the spectrum in
(a) (see text for details). Spectrum (b) contains many fewer strong
lines. Panel (c) is the same as panel (b), but assuming the same
amplitude of the ionizing background as in panel (a).}
\label{fig:spec}
\end{figure*}

In the traditional picture of the \lya-forest, strong lines (column
density $\ge 10^{15}$ cm$^{-2}$) are produced close to collapsing
halos, which typically consist of a central concentration of high
density, cooled gas which is surrounded by a hot halo of shocked
gas. The hot halo itself is collisionally ionized and consequently
contributes very little to the absorption. This is illustrated in
Fig.~\ref{fig:spec2}, where we compare a stretch of absorption spectrum
with and without the contribution from hot gas. Here, gas at a given
density $\rho$ is considered \lq hot\rq~ if its temperature is more
than 20 per cent higher than the minimum temperature at that
density. Even though this is a rather low limit, the spectrum shown in
Fig.~\ref{fig:spec2} is the only stretch of spectrum within a longer
spectrum of length $4\times 10^4$ km s$^{-1}$ (the typical length of a
$z=3$ QSO \lya~spectrum) where there is {\em any} noticeable
contribution of the hot component. Consequently, as long as feedback on
the hot component has no dynamical effects, it is unlikely to affect
the observed properties of the \lya-forest.

However, if feedback also affects the cold and dense gas in the
surroundings of a halo, it may have an important effect on the \lya~
spectrum. We have run a standard friends-of-friends group finder
(linking length 0.2 times the mean inter particle separation) on all
stars in the simulation and identified \lq galaxies\rq~ as groups of
more than ten \lq stars\rq~ (one particle has mass $1.3\times
10^6\msun$ so the minimum stellar mass in a galaxy is $1.3\times
10^7\msun$). We then assumed that galaxy feedback is able to heat all
the gas within a physical radius $R$ around the galaxy to such high
temperatures that it no longer contributes to the absorption. We choose
$R=150$\kpc (i.e., a significant fraction of the virial radius of a
large galaxy), in which case the volume filling factor, $n_{\rm
gal}(4\pi/3) R^3=1$, where $n_{\rm gal}$ is the number density of
galaxies. (The true volume filling factor of hot gas is then much
lower, since these clustered halos overlap considerably.)

Fig.~\ref{fig:spec} compares a stretch of simulated spectrum (at
$z\approx 3$, panel a), with a simulated spectrum with feedback (panel
b). The simulated spectra are scaled to have the same mean absorption
as the spectrum of QSO 1422+231 (Rauch et al. 1997). Feedback
significantly decreases the number of strong lines, since the gas that
originally produced these lines is now assumed to be hot (compare
e.g. the lines at $v\approx 2000$ km s$^{-1}$ in panels a and
b). Conversely, weak lines in panel (a) appear much stronger in panel
(b), since the spectra are scaled to the same mean absorption. Spectrum
(a), without feedback, looks similar to the observed spectrum of QSO
1422 in terms of the number of strong lines, whereas spectrum (b),
which mimics strong feedback, has far fewer strong lines. Panel (c) is
a spectrum with feedback, but not scaled to the same mean absorption,
i.e., assuming the same amplitude for the ionizing background as in
panel (a). This illustrates the effect of the rescaling.

\begin{figure}
\setlength{\unitlength}{1cm}
\centering
\begin{picture}(8,9)
\put(-1.5, -3){\includegraphics{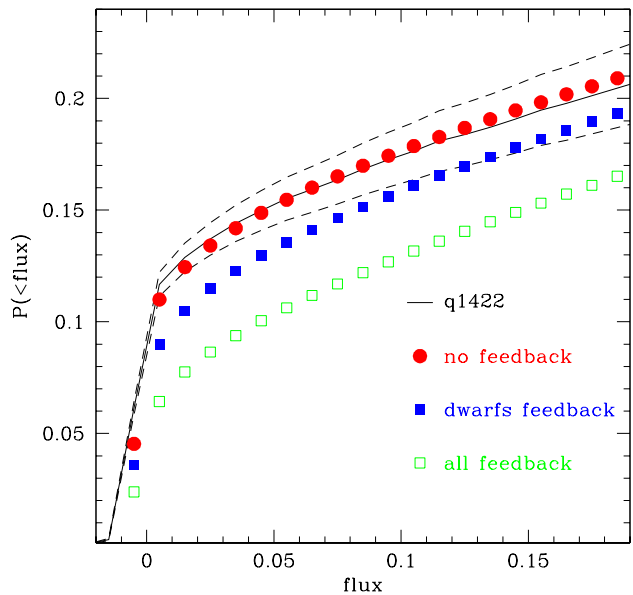}}
\end{picture}
\caption{Comparison of the cumulative fraction of pixels with flux
greater than some value. Full line: spectrum of QSO 1422, with
1$\sigma$ error contours (dashed lines); filled circles: simulated
spectrum without feedback; squares: simulated spectra with feedback due
to dwarf galaxies (filled squares) and due to all galaxies (open
squares), see text for details. Feedback has a significant effect on
the number of pixels with high absorption, flux $\le 0.2$.}
\label{fig:1pt}
\end{figure}

\begin{figure}
\setlength{\unitlength}{1cm}
\centering
\begin{picture}(8,9)
\put(-1.5, -3){\includegraphics{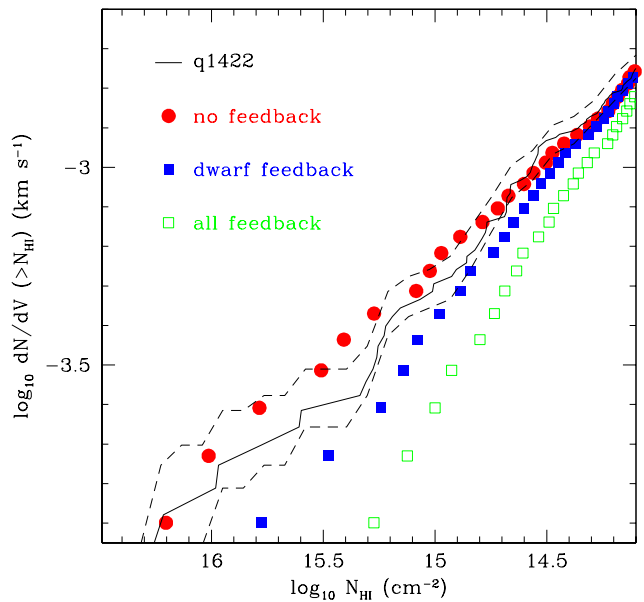}}
\end{picture}
\caption{Cumulative number of Voigt-profile lines per unit velocity (km
s$^{-1}$), with column density larger than some value. Symbols are the
same as in Fig.~\ref{fig:1pt}. The simulations that mimic feedback
(open and filled squares) produce significantly fewer lines than the
simulation without feedback (filled circles). The latter reproduces the
observations (full line) very well. Dashed lines indicate the 25 and 75 per
centiles for the observations, determined using bootstrap resampling.}
\label{fig:vpfit}
\end{figure}

Indeed, the removal of saturated regions due to feedback has a clear
observational signature in a more detailed analysis. Fig.~\ref{fig:1pt}
compares the one-point functions (histogram of the flux per pixel) of
the spectra shown earlier. Whereas the simulated spectrum without
feedback reproduces the observed one-point function very well, the
simulations that mimic feedback produce significantly fewer pixels with
high absorption (flux $\le 0.2$). Two such simulations are illustrated
in Fig.\ref{fig:1pt}, the one labelled \lq dwarfs\rq~ includes feedback
from galaxies with mass $1.3\times 10^7\le M/\msun\le 1.3\times 10^8$,
the one labelled \lq all\rq~ includes feedback from all galaxies above
$1.3\times 10^7\msun$.  In both cases, the radius of the sphere $R$
around the galaxy from which the cold gas is removed, prior to
computing the mock absorption spectra, equals $R=150$\kpc, and the
volume filling factors are then $\approx$ 0.1 and $\approx 1$, for
dwarfs and all galaxies, respectively. [We recall that these volume
filling factors are simply the ratio of the sum of the volumes of all
bubbles to the total volume. The {\em true} fractions of volume
encompassed by the bubbles, i.e. taking into account that the bubbles
overlap, are 6.5 and 30 per cent, for the dwarfs and all galaxies
respectively. If we increase $R$ to 220\kpc~ for the dwarfs simulation
-- in which case the volume filling factor increases to $\sim 1$ (true
volume filling factor 38 per cent), the one-point function (as well as
the number of strong lines, to be discussed next), hardly changes.]
Cold gas surrounding the more massive galaxies produces a relatively
large fraction of the saturated pixels, which explains the difference
between imposing feedback on all galaxies, or only on the small ones.

Cen (1997) argued that the high-end of the \lya~optical depth
distribution is quite sensitive to the amplitude of the power spectrum,
because it probes the tail of the (e.g. Gaussian) density probability
distribution. If feedback effects are strong then the ability of this
statistic to distinguish between cosmological models may be partly
compromised.

The consequence of having fewer strong lines, in the presence of
feedback, can also be seen when plotting the column density
distribution, $d^2N/d\NHI/dV$, the fraction of lines per unit column
density, per unit spectral length in km s$^{-1}$. We have fitted Voigt
profiles to both simulations and to the observed spectrum of QSO 1422,
using the same fitting algorithm (an automated version of \vpfit; Webb
1987, Carswell et al. 1987). Fig.~\ref{fig:vpfit} shows that the
simulation without feedback reproduces the observed numbers of strong
lines very well, whereas the simulations that mimic feedback
significantly under produce the number of strong lines. Note that the
shape of the column-density distribution is sensitive to the amount of
large-scale power (e.g. Gnedin 1998). This could be used to distinguish
the effects of feedback from those of large-scale power, on the
column-density distribution.

\begin{figure*}
\setlength{\unitlength}{1cm}
\centering
\begin{picture}(15,12)
%\put(-1.5, -12.){\special{psfile=nh16-bw.ps hscale=100 vscale=100}}
\put(-1.5, -8.){\includegraphics{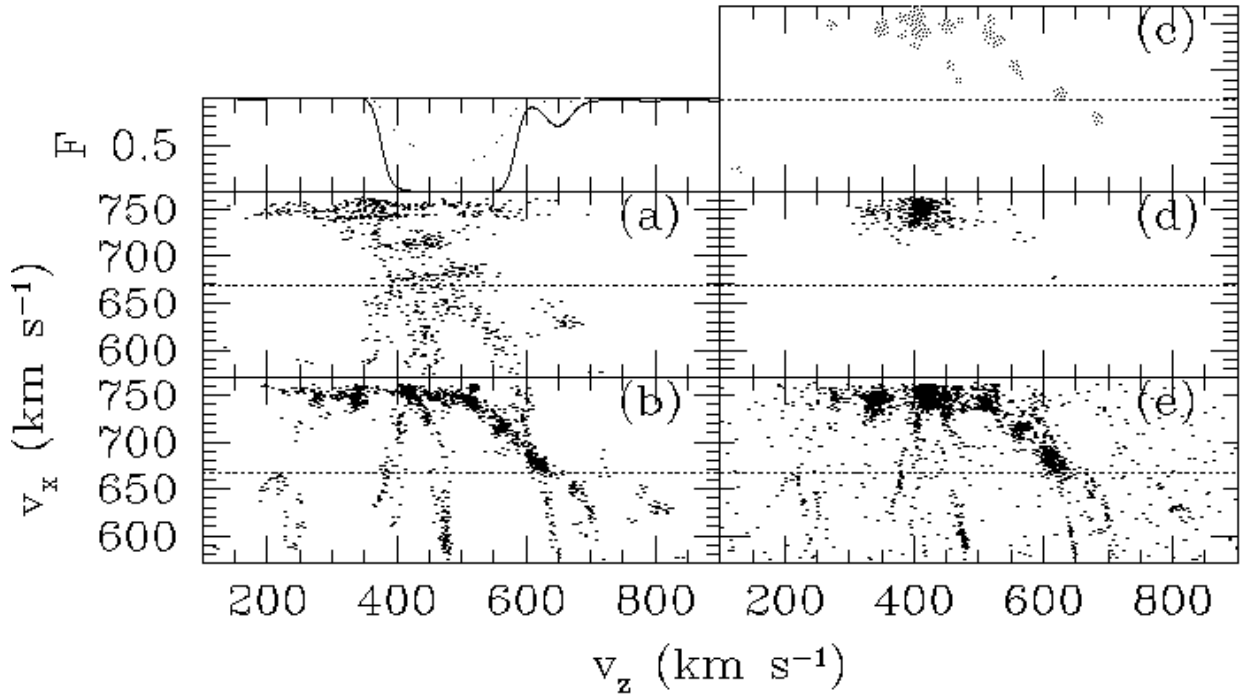}}
\end{picture}
\caption{Geometry of the structure that produces a strong \lya~
line. Top left panel: \lya~absorption spectrum (flux versus velocity)
of a line with column density $1.4\times 10^{16}$ cm$^{-2}$ (full line)
and the corresponding \lyb~profile (dashed). Panels a, b and d show the
gas distribution in a thin slice along the sight-line (the dashed
horizontal line in panels a-e) that produces this strong line. Plotted
are gas particles with $\rho\ge\langle\rho\rangle$ and temperature
$T\le 2.5\times 10^4$K (panels a-b) and hot gas particles with $T\ge
2.5\times 10^4$K (panel d). Positions are labeled in Hubble velocity
$v=H(z)r$. The slice in the $y$-direction (perpendicular to the page)
is 10 km s$^{-1}$ (20 km s$^{-1}$ for panel d). Star particles are
plotted as stars in panel (c). In panel (a), $v_z$ refers to the
redshift space velocity, $Hr+v_{\rm pec}$. Panel (e) plots the
corresponding dark matter particles.  The absorption line is a
superposition of three to four filaments which overlap in redshift
space due to the high peculiar velocities induced by the nearby large
mass concentration. These high velocities make it difficult to identify
the structure in redshift space (panel a) with the corresponding
filaments in real space (panel b).  The massive system at the top is
embedded in hot, shocked gas, but there is very little hot gas
associated with the filaments that produce the line. Note that the gas
density is an interpolation over the plotted SPH particles, so that
particles not directly hit by the plotted line-of-sight may still
contribute to the absorption.}
\label{fig:nh16}
\end{figure*}

The above considerations show that this type of maximal feedback,
occurring in a spherical region around the forming halo can be ruled
out. However, the distribution of gas that gives rise to absorption
lines is actually far from spherical. In fig.~\ref{fig:nh16} we plot
the matter distribution of a system that produces a strong line with
column density $1.4\times 10^{16}$ cm$^{-2}$. This line is a
superposition of the absorption from three to four different filaments,
which blend together in redshift space due to peculiar velocity
gradients. The distribution of the cold gas which is responsible for
the line is filamentary. Presumably, a galactic outflow from the stars
near the dense knot at $v_z\sim 600\kms$, which contributes most to the
absorption, would preferentially occur perpendicular to the filament
along the path of least resistance, in which case the line may not
suffer strongly from feedback after all. But the consequence of
directing all the feedback momentum into the low-density surroundings,
is that the accretion rate onto the halo, which is dominated by infall
along the filament, cannot be strongly affected by the feedback. That
is, if feedback is able to limit the accretion rate onto dwarf halos,
then it will necessarily also influence the properties of strong \lya~
absorption lines. Note that the individual components that give rise to
the strong line can be recognized in the higher-order \lyb~line, if it
is not saturated. Therefore, higher order Lyman-lines should be good
probes of the state of the cold gas in the surroundings of forming
halos, and can be used to examine whether feedback is able to
significantly stir the accreting gas.

Clearly, a more detailed analysis, for example based on probing
substructure in strong lines using higher order transitions, is
required, but it does seem safe to conclude that the properties of
strong lines may be a valuable probe of feedback round forming
galaxies. A detailed comparison of simulations without feedback with
high-resolution observations may shed light on whether galaxy feedback
is an important factor in determining the properties of strong lines.

\subsection{Winds from dwarf galaxies}
\begin{figure}
\setlength{\unitlength}{1cm}
\centering
\begin{picture}(8,11)
\put(-1.5, -3){\includegraphics{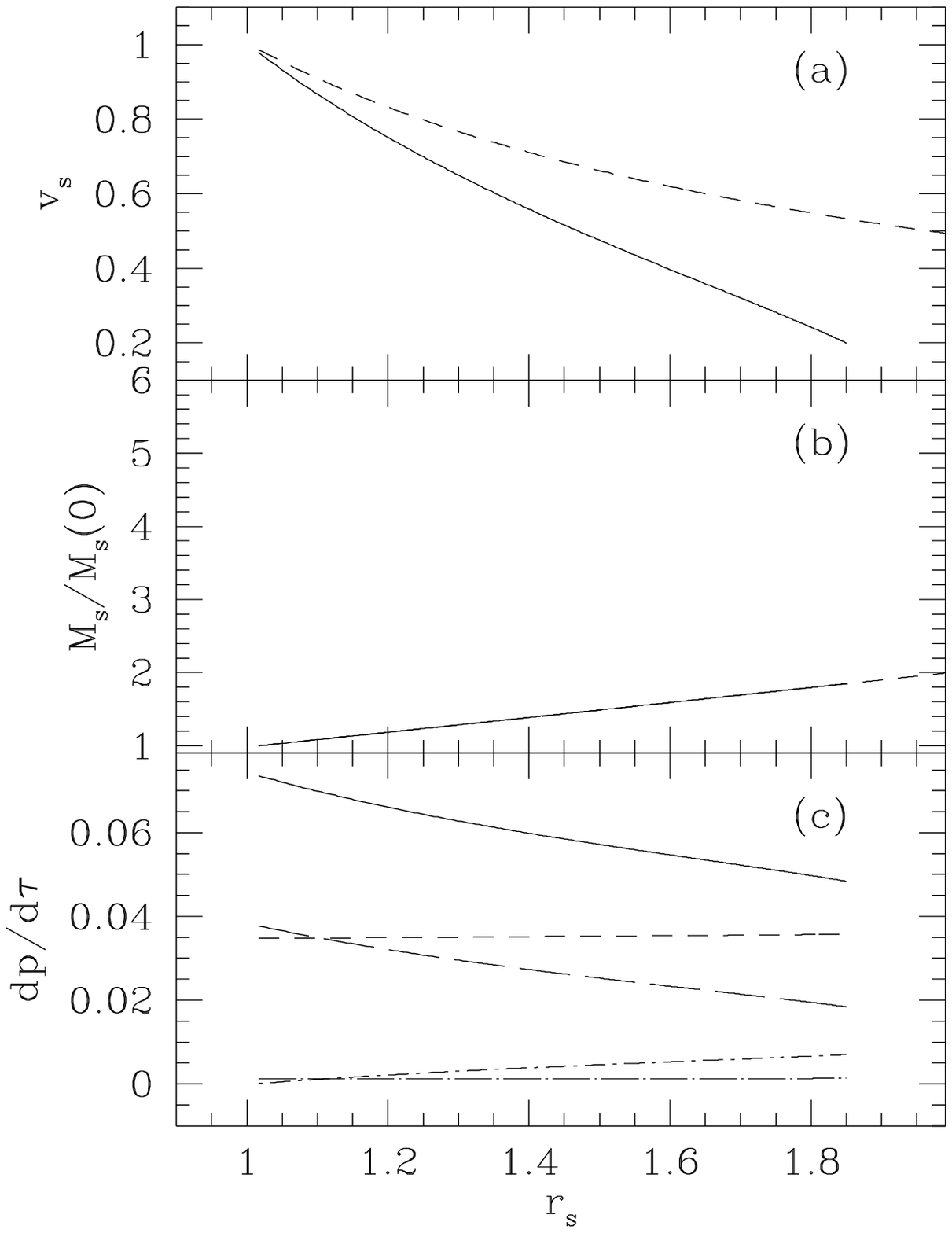}}
\end{picture}
\caption{Dynamics of a wind with mass-loss rate $\dot M_w=0.5
M_\odot$/yr and wind-speed 100\kms, emanating from a galaxy with
circular velocity 50\kms into the IGM with sound speed 20\kms at a
redshift $z=3$.  (We assumed in addition $h=0.65$, $\Omega_b=0.05$ and
$\Omega_m=0.3$ for the cosmological parameters.)  Panel (a) depicts the
velocity of the shell (in units of the wind velocity) versus its
position (in units of the virial radius $R_{200}=17.5$\kpc~ of the
halo). Panel (b) shows the mass of the shell, in units of the initial
mass. The dashed curves neglect the forces acting on the shell. Panel
(c) shows the magnitude of the dimensionless forces acting on the
shell. From top to bottom, the curves denote the total force (full
line), gravity (short dashed), ram-pressure force of the surrounding
IGM (long dashed), ram-pressure of the wind (dot-short dashed), and the
thermal pressure of the IGM (short dashed-long dashed.}
\label{fig:windacc}
\end{figure}

\begin{figure}
\setlength{\unitlength}{1cm}
\centering
\begin{picture}(8,8)
\put(-1.5, -3){\includegraphics{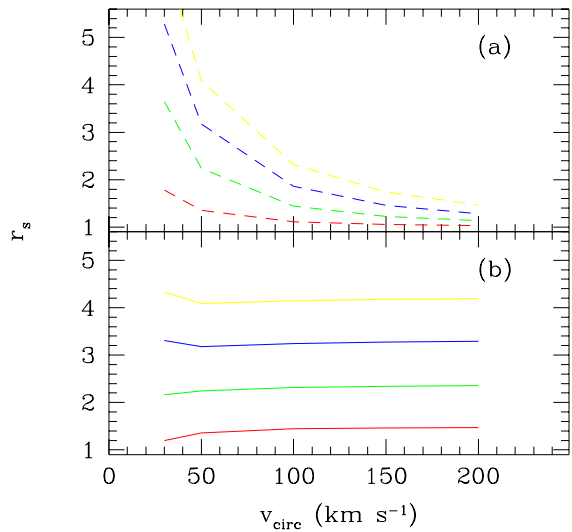}}
\end{picture}
\caption{Stagnation radius of the shell (in units of the virial radius
$R_{200}$ of the halo) versus circular velocity of the halo, for wind
velocities of 50, 100, 150 and 200\kms (panel a), and 1, 2, 3, 4
$\times V_{\rm circ}$ (panel b).}
\label{fig:rs}
\end{figure}

In the previous section we discussed the effects of feedback on the
immediate surroundings of the galaxy. We showed that feedback cannot
influence all the cold gas around forming halos out to $\sim 150$\kpc~
without significantly affecting the properties of strong lines as
well. Here we discuss in more detail whether outflows from dwarf
galaxies may themselves be responsible for producing absorption lines.

The mass-to-light ratio in dwarf galaxies is typically of the order
$M/L\sim 100M_\odot/L_\odot$ (e.g., Carignan \& Freeman 1988, Lake,
Schommer \& van Gorkom 1990), much higher than in spirals or
ellipticals. Presumably, a wind associated with feedback from
star-formation was able to blow most of the gas out of the shallow
potential well of the galaxy. Indeed, in the model of Efstathiou
(2000), a galactic wind is able to remove 60-80 per cent of the gas
contained in a dwarf galaxy, over a time-scale of order 1Gyr. In these
models, the outflow velocity at the virial radius is typically of order
a few times the escape velocity, which in turn is typically a few times
the circular speed ($\vcirc$) of the halo. In this case, the wind may
evacuate a cavity in the IGM, surrounded by a shell.

The momentum conservation equation for a spherical shell with mass
$M_s$ and velocity $V_s$, which is being blown into the surrounding IGM
by a galactic wind with mass-loss rate $\dot M_w$ and velocity $V_w$ is
\begin{eqnarray}
{d\over dt} \left( M_s V_s\right)\!\!\!\!\!\! &=& \dot M_w V_w (1-{V_s\over V_w}) - 4\pi R_s^2\rho
\left\{{C^2/\gamma} + V (V_s+V)\right\}\nonumber\\ 
&-& {G M_g M_s\over R_s^2}\,,
\label{eq:momentum}
\end{eqnarray}
where $R_s$ is the current position of the shell. The first term is the
momentum accreted from the wind. The second term is the deceleration of
the shell due to the pressure of the surrounding IGM (density
$\rho(R_s)$, sound speed $C$ and adiabatic index $\gamma$), and due to
the sweeping-up of matter which rains onto the expanding shell with
velocity $V$. Finally, the last term is the gravitational deceleration
due to the mass $M_g$ interior to the shell ($G$ denotes Newton's
constant of gravity). The mass accretion rate of the shell due to
accretion from the wind and the sweeping-up of the IGM is
\begin{equation}
{d\over dt} M_s= \dot M_w (1-{V_s\over V_w}) + 4\pi R_s^2\rho
(V_s+V)\,.
\label{eq:mass}
\end{equation}

The initial conditions can be set at the virial radius $R_{200}$ of the
halo, defined as usual as the radius within which the mean overdensity
is 200 times the critical density. Mo, Mao \& White (1998) give the
relation between the virial radius, the enclosed mass $M_{200}$, the
circular velocity $V_{\rm circ}$ and the Hubble constant at redshift
$z$ as
\begin{equation}
R_{200} = {V_{\rm circ}\over 10 H(z)}\,,
\label{eq:r200_def}
\end{equation}
\begin{equation}
M_{200} = {V_{\rm circ}^3\over 10GH(z)}\,.
\label{eq:m200_def}
\end{equation}
We will assume $V_s=V_w$ at $R_s=R_{200}$ and denote the fraction of
(baryonic) mass swept-up by the wind by the time it reaches the virial
radius by $\alpha\le 1$. We also assume the wind-speed to be
constant. Finally we need to specify the density profile of the
material surrounding the galaxy, for which we assume
\begin{equation}
\rho(R) = \left[ {200\over 3} \left({R_{200}\over R}\right)^2 + 1\right]\rho_c\,,
\end{equation}
where $\rho_c=3H^2/8\pi G$ is the critical density. The gas density is
$\Omega_b/\Omega_m\rho$. These equations are valid at high redshift
where $\langle\rho\rangle\approx\rho_c$.

Equations~(\ref{eq:momentum}-\ref{eq:mass}) can be cast in
dimensionless form, using $R_{200}$, $V_w$ and $M_{200}$ as the units
of distance, velocity and mass, respectively. Using lower-case letters
to denote dimensionless variables we find
\begin{eqnarray}
{d\over d\tau} \left(m_s v_s\right)\!\!\!\!\!\! &=& {\dot M_w t_\star\over M_{200}} (1-v_s) 
\nonumber\\
&-&\Omega_b r_s^2 (r_s^{-2}+3/200) \left[ (v+v_s)v + c^2/\gamma\right]\nonumber\\
&-&v_{\rm circ}^2 {(1+\chi-m_s) m_s/r_s^2}\\
{d\over d\tau} m_s\!\!\!\!\!\! &=& {\dot M_w t_\star\over M_{200}} (1-v_s) +
\Omega_b r_s^2(r_s^{-2}+3/200)(v+v_s)\,\nonumber\\
\end{eqnarray}
where $\chi\equiv (r_s-1) + (r_s^3-1)/200$ and the time
variable $\tau=t/t_\star$ with $t_\star=R_{200}/V_w$. 

Fig.~\ref{fig:windacc} illustrates the dynamics of a wind emanating
from a small galaxy with circular velocity 50\kms. We assumed that
feedback is very efficient and the initial shell mass $M_s=\alpha
(\Omega_b/\Omega_m) M_{200}$ with $\alpha=1$. We further assumed that
gas is accreting onto the galaxy with velocity $V=V_{\rm circ}$ and
integrated the equations until the stagnation point where the velocity
becomes sonic, $V_s=C$. The deceleration of the shell is dominated by
gravity (mostly from the halo) in this case, with a non-negligible
contribution from the ram-pressure of the infalling gas. Nath \&
Trentham (1997) followed Tegmark, Silk \& Evrard (1993) and neglected
the gravity of the halo, as did Ferrara, Pettini \& Shchekinov (2000)
when they estimated the distance to which a wind-blown shell will
travel and pollute the IGM. In our case, the shell travels out to $\sim
1.8$ times the viral radius, thereby doubling its mass. The dashed
lines in panels (a) and (b) illustrate the case where forces are
neglected, in which case the shell reaches a considerably larger
distance, sweeping-up $\sim V_w/C$ times its initial mass.

Fig.~\ref{fig:rs} illustrates the stagnation point of the shell for
different wind velocities, as a function of the circular velocity of
the halo. We have assumed that a wind with mass-loss rate $\dot
M_w=0.5M_\odot$ yr$^{-1}$ blows for 0.5Gyr (so that it can in principle
remove a substantial fractrion of the baryons over a long time; see
also Efstathiou 2000), and plot the position of the shell when it
stagnates, $V_s=C$, or until $t=1.0$Gyr. In small halos, winds of
$100-150$ \kms are able to expell gas up to 3-6 times the virial
radius, but the increased gravity in more massive halos prevents the
shell from traveling far. On the other hand, if the wind velocity is a
given fraction of the circular speed, then slow winds blow further in
more massive halos, since such winds can blow for longer before the
shell speed becomes equal to the sound speed of the IGM.

The virial radius $R_{200}$ of a halo is related to its circular
velocity by (Eqs.~\ref{eq:r200_def}-\ref{eq:m200_def})
\begin{equation}
R_{200} = 17\kpc\,{V_{\rm circ}\over 50 \kms}
\left({\Omega_m\over 0.3}\right)^{-1/2}
\left({1+z\over 4}\right)^{-3/2}\,,
\label{eq:r200}
\end{equation}
and the corresponding virial mass 
\begin{eqnarray}
M_{200}&=&1\times 10^{10} \left({V_{\rm circ}\over 50\kms}\right)^3
\left({\Omega_m\over 0.3}\right)^{-1/2}\nonumber\\
&\times &\left({h\over 0.65}\right)^{-1}
\left({1+z\over 4}\right)^{-3/2}\msun\,.
\label{eq:m200}
\end{eqnarray}
Therefore for the wind parameters in Fig.~(\ref{fig:windacc}) the final
shell radius $R_s\sim 30$\kpc~ and the mass in the shell $M_s\sim 3.3\times
10^9\msun$, where we assumed that $\Omega_b/\Omega_m=0.05/0.3=0.17$.

The cross-section of shells of such a size is quite high, with a mean
number of intersections per unit redshift of
\begin{eqnarray}
\label{eq:dNdz}
{dN\over dz} &\approx& 50 {n_0\over 0.5\,(h^{-1}{\rm Mpc})^3} \left({R_s\over
30{\rm kpc}}\right)^2\left({h\over 0.65}\right)^{-1}\nonumber\\
&\times & \left({\Omega_m\over 0.3}\right)^{-1/2} \left({1+z\over 4}\right)^{1/2}\,,
\end{eqnarray}
where $n_0\sim 0.5(h^{-1}{\rm Mpc})^{-3}$ is the co-moving
number-density of dwarfs (as determined from our galaxy-cluster
normalized simulation). If such a shell produces absorption lines, then
a significant fraction of all \lya~lines could be due to such
shells. (A $z=3$ quasar spectrum contains typically of order 300 lines
above a column density of $10^{13}$ cm$^{-2}$ per unit redshift).

Note that one would expect such wind blown bubbles to produce {\em two}
lines, when punctured by a sight-line, separated by $\sim
100$\kms. Cristiani et al (1997) analyzed \lya-lines of several
high-resolution quasar spectra and concluded that the two-point
correlation function of the lines shows clustering at such scales. In
contrast, Pettini et al (1990) and Stengler-Larrea \& Webb (1993)
claimed a non-detection of such a signal. Part of the confusion might
be caused by the non-uniqueness of Voigt-profile fits, especially in
the presence of severe line-blending occuring at high redshifts. A
uniform comparison of high-resolution spectra with simulations is
required to check for the reality of such a signal.

Below we will estimate the column density of such a shell, but first we
would like to point out that there are significant uncertainties in the
estimate of the shell's radius. Dwarf galaxies in simulations are
invariably embedded in filaments at over densities of a few, which have
significant non-spherical infall velocities which will oppose
outflows. Therefore a more realistic description of the shell's motion
should take into account such a filamentary matter distribution.

More importantly, it is not clear how efficiently the outflowing gas
will couple to the surrounding IGM. The wind may break-up into clouds
which coast to large distances without sweeping-up a significant
fraction of the ambient IGM. In that case, although some of the
outflowing gas may still reach large distances, the swept-up mass may
be much less than that estimated above. If we assume that only a
fraction $f$ of the surrounding IGM is swept-up, with the rest in cold
clouds, then the total column density of the remaining gas
\begin{equation}
N_H \sim 3.5\times10^{19} f {M_s\over 3\times 10^9\msun}
\left({R_s\over 30\kpc}\right)^{-2}\,.
\end{equation}
Since the wind velocity ($\sim 100\kms$) is much larger than the sound
speed of the IGM, the wind results in a strong shock as it runs into
its surroundings.  In the adiabatic phase, the density of the shocked
shell is then about 4 times that of the ambient medium ($n_1$), while
the temperature is about $T_2 \sim 5\times 10^{5}(v_{\rm
w}/200\kms)^2{\,\rm K}$. Gas of such temperature and density, and with
a metallicity $Z> 0.01 Z_{\odot}$, has a cooling time shorter than its
age [see equation (\ref{coolingtime}). At these temperatures, cooling
will be dominated by metals, hence $\Lambda_{23}\gg 1$.].  Thus, as the
shock propagates into the medium and sweeps-up gas, it can cool down to
$\sim 10^4\,{\rm K}$ and become dense, with a density $n_3\sim {\cal
M}_1^2 n_1\sim 100 n_1$ ($\sim 10^{-3}{\,\rm cm}^{-3}$ at $z\sim 3$),
where ${\cal M}_1$ is the initial Mach number of the shock. Under these
assumptions, the column density of the shell is
\begin{eqnarray}
\NHI &\sim& 1.4\times 10^{16}f\,\left({T_4^{-3/4}\zeta\over
J_{21}}\right) \left({n_3\over 10^{-3} {\,\rm cm}^{-3}}\right)\nonumber\\
&\times &{N_H\over 3.5\times 10^{19} {\rm cm}^{-2}}\,,
\label{eq:nhi}
\end{eqnarray}
where we used equation~(\ref{eq:peq}) for converting the total to the
neutral column density.

The physical parameters for the winds from a dwarf versus those from a
LBG, discussed in Section~\ref{sect:lbg}, are quite different. For the
assumed mass-loss rates (0.5 for the dwarf versus 100$M_\odot$
yr$^{-1}$ for the LBG) and wind velocities (100 versus 200\kms) for
these systems, the wind densities differ by a factor
1/100. Consequently for the LBG, it is the dense, cold wind that
produces most of the absorption, since the swept-up hot shell cannot
cool. The reverse is true for the wind from the dwarf galaxy, for which
the tenuous wind produces no significant absorption, but the dense
swept-up shell produces lines.

If the wind interacts strongly with the IGM, $f\sim 1$, the high column
density of the lines given by equation~(\ref{eq:nhi}) combined with the
rate of incidence of the shells, $dN/dz\sim 50$, predicts of order 100
lines with columns larger than $10^{16}$~cm$^{-2}$ per quasar spectrum,
which is already much more than the typical total number $\sim 20$ of
such lines in an observed spectrum. Therefore, these considerations
require $f\ll 1$, that is the wind cannot retain its integrity but must
break-up into clouds. These clouds, which may have higher
column-densities, are then not subject to the momentum conservation
Eq.~({\ref{eq:momentum}) and can coast to large distances $r\sim
0.3$~\Mpc~ if they move at $100\kms$ over the Hubble time,
Eq.~(\ref{eq:thub}). Since these clouds are polluted by metals of the
SNe explosions that drive them out of the galaxy in the first place,
such winds should be able to deposit heavy elements throughout the
IGM. Clearly a much more detailed model is required to compute the
covering factor of such clouds. Alternatively, the matter density in
the surroundings of the halo may limit the expansion of the shell much
more than we have computed here, for example because the infall
velocities are higher than we have assumed, or because the distribution
is non-spherical and the winds are funneled into the low-density
surroundings without sweeping-up much matter.

Finally we would like to make a remark on the effects of such winds on
neighbouring halos. In numerical simulations of hierarchical
clustering, each dark matter halo always contains a swarm of smaller
halos down to the resolution scale (e.g. Moore et al 1999). A galactic
wind would presumably remove most of the gas from such sub-clumps,
leaving these sub-structures mostly dark.

In summary: strong \lya~lines may be excellent probes of feedback
effects on galactic scales. If galactic feedback is assumed to heat a
large fraction of the cold gas in the surroundings of forming halos,
then many fewer strong lines are produced than are observed in QSO
spectra. Shells formed from gas swept-up by galactic winds from dwarf
galaxies would over-produce the number of strong absorption
lines. Therefore we argued that these winds should break-up into
clouds, which can then coast to large distances and pollute the IGM
with metals, or be funneled into the low-density surroundings without
producing a strong line.

\section{Summary}
The paradigm in which the \lya-forest is produced by the low-density
photo-ionized IGM, which traces the structure of the underlying
dark-matter perturbations, has been very successful in explaining
properties of observed high-resolution quasar spectra. However, models
of galaxy and galaxy cluster formation require large amounts of
feedback in order to match the observations. We have investigated
possible observational signatures of this type of feedback in
\lya~spectra.

Pre-heating is often invoked to explain the observed X-ray properties
of galaxy clusters. A QSO sight-line intersects of order one such
proto-cluster per unit redshift. If the intra-cluster gas is hot, it
will produce a large gap with little absorption in the spectrum, which
could be detected at high redshift, $z\sim 3-4$, when the \lya-forest
opacity is high. If the proto-cluster gas is in a two-phase medium, the
gas in pressure-confined cold clouds could produce a single strong line
with associated metals. Such a model can be tested observationally, by
probing the sub-structure of such absorbers using e.g. higher-order
Lyman lines, or by using multiple sight-lines. Dense groups of
Lyman-break galaxies at $z\sim 3-4$ are likely sites of proto-clusters,
and QSO spectra taken through such a structure could be used to put
limits on cluster pre-heating. Winds associated with star formation in
Lyman-break galaxies will also produce lines with column-densities
$> 10^{15}$ cm$^{-2}$, plausibly associated with metals escaping
from the galaxy. Such lines can be used to study the feedback process
in these galaxies.

Feedback on galactic scale will influence the properties of strong
$\NHI\sim 10^{16}$ cm$^{-2}$ \lya~lines, since these are produced close
to forming halos. If feedback is able to heat a significant fraction
of the cold gas in the surroundings of a galaxy, then many fewer strong
lines will be produced. However, the outflows may be preferentially
directed into the low-density surroundings, in which case strong lines
may survive. Shells associated with winds from dwarf galaxies 
will over produce the number of absorption lines, unless the wind
breaks-up into clouds. These clouds can coast to large distances and
pollute the IGM with metals. Alternatively, the matter surrounding the
galaxy may funnel the wind into the low-density IGM, without producing
1strong lines.

These simple arguments suggest that feedback may have an observable
effect on the \lya-forest. Some of our predictions on cluster feedback
can be tested by correlating \lya-absorption spectra with the
Lyman-break galaxies along the line-of-sight. Simulations of dwarf
galaxy formation that include a wind, can be used to examine the
interaction of a galactic wind with accreting gas.

\section*{Acknowledgements}
We thank M. Rauch and W. Sargent for providing us with the spectrum of
QSO 1422+231, and G. Efstathiou for many discussions and suggestions. T
T acknowledges partial funding from PPARC and J S thanks the Isaac
Newton trust and PPARC for the award of a studentship. This work has
been supported by the \lq Formation and Evolution of Galaxies\rq~
network set up by the European Commission under contract ERB
FMRX-CT96086 of its TMR programme. Research conducted in cooperation
with Silicon Graphics/Cray Research utilizing the Origin 2000
supercomputer at the Department for Applied Mathematics and Theoretical
Physics (DAMTP), Cambridge.


\begin{thebibliography}{}
%\bibitem[] {} Abel T., Haehnelt M.G., 1999, ApJL, 520, 13
\bibitem[\protect\citename{}]{} Adelberger K.L., Steidel C.C., Giavalisco M., Dickinson M.,
Pettini M., Kellogg M., 1998, ApJ, 505, 18
\bibitem[\protect\citename{}]{} Bahcall J.N., Salpeter E.E., 1965, ApJ,
142, 1677
\bibitem[\protect\citename{}]{} Balogh M.L., Babul A., Patton D.R.,
1999, MNRAS, 307, 463
\bibitem[\protect\citename{}]{} Baugh C.M., Cole S., Frenk C.S.,
Lacey C.G., 1998, ApJ, 498, 504
\bibitem[Bi, Boerner and Chu (1991)]{1991A&A...247..276B} Bi H. G., 
Boerner G., Chu Y. 1991, \aap, 247, 276 
\bibitem[Bi and Davidsen (1997)]{1997ApJ...479..523B} Bi H., Davidsen
A F,1997, ApJ, 479, 523 
\bibitem[]{} Breitschwerdt D. McKenzie J.F., V\"olk H.J., 1991,
A\&A~, 245, 79
\bibitem[\protect\citename{}]{} Burles S., Tytler D., 1998, 
ApJ, 499, 699

\bibitem[Carignan and Freeman (1988)]{1988ApJ...332L..33C} Carignan C,
Freeman K C, 1988, ApJ, 332, L33 


\bibitem[\protect\citename{Carswell \etal 1987}]{carswellwba87} 
Carswell R.F., Webb J.K., Baldwin J.A., Atwood B., 1987, ApJ, 319,
709
\bibitem[\protect\citename{}]{} Cavaliere A., Menci N., Tozzi P.,
1997, ApJ, 484, 21
\bibitem[] {} Cen R., 1997, ApJL, 479, 85
\bibitem[\protect\citename{Cen \etal 1994}]{cenor94} Cen R.,
Miralda-Escud\'e J., Ostriker J.P., Rauch M., 1994, ApJL, 437, L9

\bibitem[\protect\citename{}]{} Cowie L.L., Songaila A., Kim T.-S.,
Hu E.M., 1995, AJ, 109, 1522
\bibitem[] {} Cristiani S., D'Odorico S., D'Odorico V., Fontana A.,
Giallongo E., Savaglio S., 1997, MNRAS, 285, 209
\bibitem[\protect\citename{}]{} David L.P., Forman W., Jones C.,
1991, ApJ, 380, 39
\bibitem[\protect\citename{}]{} Dekel, A., Silk, J., 1986, ApJ, 303, 39
\bibitem[Efstathiou (1992)]{1992MNRAS.256P..43E} Efstathiou G. 1992, MNRAS, 256, 43P
\bibitem[\protect\citename{}]{} Efstathiou G., 2000, MNRAS, 317, 697
\bibitem[\protect\citename{}]{} Efstathiou G., Bridle S.L., Lasenby
A.N., Hobson M.P., Ellis R.S., 1999, MNRAS, 303, 47
\bibitem[] {} Efstathiou G., Schaye J., Theuns T, 2000, Philos. Trans. R. Soc. Lond. A, 358, 2049

\bibitem[\protect\citename{}]{} Eke V.R., Cole S., Frenk C.S.,
1996, MNRAS, 282, 263
\bibitem[\protect\citename{}]{} Ellison S.L., Songaila A., Schaye J.,
Pettini M., 2000, AJ, 120, 1175
\bibitem[\protect\citename{}]{} Evrard A. E., Henry J. P., 1991, ApJ,
383, 95
\bibitem[\protect\citename{}]{} Ferrara A., Pettini M., Shchekinov Y.,
2000, MNRAS, 319, 539
\bibitem[\protect\citename{}]{} Freedman J.B., Mould J.R., Kennicutt
R.C., Madore B.F., 1999, in IAU Symposium 183, Cosmological parameters
and the evolution of the universe, ed. K. Sato (Kluwer:Dordrecht)
\bibitem[\protect\citename{}]{} Giavalisco M., Steidel C.C., Adelberger K.L., Disckinson M.E.,
Pettini M., Kellogg M., 1998, ApJ, 503, 543
\bibitem[\protect\citename{}]{} Gnedin N., 1998, MNRAS 299, 392
\bibitem[\protect\citename{}]{} Governato  F.,Baugh C. M., Frenk
C. S., Cole S.,  Lacey C. G., Quinn T., Stadel J., 1998, Nature,
392, 359
\bibitem[\protect\citename{}]{} Gunn J.E., Peterson B.A., 1965, ApJ,
142, 1633
\bibitem[\protect\citename{}]{} Heckman T., 2000, To appear in the
Philosophical Transactions of The Royal Society, Series A', (astro-ph/9912029)
\bibitem[\protect\citename{}]{} Heckman T.M., Armus L., Miley G.K.,
1990, ApJS, 74, 833
\bibitem[\protect\citename{}]{} Heckman T.M., Lehnert M.D.,
Strickland D.K., Armus L., 2000, ApJS, 129, 493
\bibitem[\protect\citename{Hernquist \etal 96}]{Hernquistetal96}
Hernquist L., Katz N., Weinberg D.H., Miralda-Escud\'e J., 1996,
ApJL, 457, L51
\bibitem[\protect\citename{}]{} Jones C., Forman W., 1984, ApJ, 276,
38
\bibitem[\protect\citename{}]{} Kaiser N., 1986, MNRAS, 219, 785
\bibitem[\protect\citename{}]{} Kaiser N., 1991, ApJ, 383, 104
\bibitem[\protect\citename{}]{} Kauffmann G., White S.D.M., Guiderdoni B.,
1993, MNRAS, 264, 201
\bibitem[\protect\citename{}]{} Kauffmann G., Colberg J.M., Diaferio
A., White S.D.M., 1999, MNRAS, 307, 529
\bibitem[\protect\citename{}]{} Lacey C.G., Guiderdoni B.,
Rocca-Volmerange B., Silk J., 1993, ApJ, 402, 15
\bibitem[Lake, Schommer and van Gorkom (1990)]{1990AJ.....99..547L}
Lake G, Schommer R A, van Gorkom J H, 1990, AJ, 99, 547 
\bibitem[\protect\citename{}]{} Larson R.B., 1974, MNRAS, 169, 229
\bibitem[\protect\citename{}]{} Lloyd-Davies E.J., Ponman T.J.,
Cannon D.B., 2000, MNRAS, 315, 689
\bibitem[McGill (1990)]{1990MNRAS.242..544M} McGill, C. 1990, \mnras, 242, 
544
\bibitem[\protect\citename{}]{} McKee C.F., Ostriker J.P., 1977, ApJ,
218, 148
\bibitem[\protect\citename{}]{} Mo H.J., 1994, MNRAS, 269, L49
\bibitem[\protect\citename{}]{} Mo H.J., Fukugita M., 1996, ApJL,
467, 9
\bibitem[\protect\citename{}]{} Mo H.J., Miralda-Escud\'e J., 1996,
ApJ, 469, 589

\bibitem[Mo, Mao and White (1998)]{1998MNRAS.295..319M} Mo H. J., Mao S. 
and White S. D. M. 1998, \mnras, 295, 319 

\bibitem[\protect\citename{}]{} Mo H.J., Mao S., White S.D.M.,
1999, MNRAS, 304, 175
\bibitem[\protect\citename{}]{} Mo H.J., Morris S.L., 1994, MNRAS,
269, 52
\bibitem[Moore, et al. (1999)]{1999ApJ...524L..19M} Moore B. , Ghigna S. 
 Governato F. , Lake G. , Quinn T. , Stadel J., Tozzi P.  1999, 
ApJL, 524, 19 

\bibitem[Mushotzky and Loewenstein (1997)]{1997ApJ...481L..63M} Mushotzky 
R. F., Loewenstein M. 1997, \apjl, 481, L63 

\bibitem[\protect\citename{}]{} Nath B.B., Trentham N., 1997, MNRAS,
291, 505
\bibitem[\protect\citename{}]{} Navarro J.F., White S.D.M., 1994,
MNRAS, 267, 401
\bibitem[\protect\citename{}]{} Navarro J.F., Frenk C.S., White
S.D.M., 1995, MNRAS, 275, 720
\bibitem[\protect\citename{}]{} Navarro J.F., Steinmetz M., 1997,
ApJ, 478, 13
\bibitem[\protect\citename{}]{} Ostriker J.P., Ikeuchi S., 1983,
ApJL, 268, 63
\bibitem[] {} Pagel B.E.J., 1999, in Galaxies in the Young Universe II,
Edt. H. Hippelein, Springer-Verlag (astro-ph/9911204)
\bibitem[] {} Peebles P.J.E., 1993, Principles of Physical Cosmology,
Princeton University Press, Princeton
\bibitem[\protect\citename{}]{} Pen U-L., 1999, ApJL, 510, 1
\bibitem[\protect\citename{}]{} Pettini M., 1999, in Chemical Evolution from Zero to High Redshift,
ed. J. Walsh \& M. Rosa (Berlin:Springer), p. 233
\bibitem[Pettini, Hunstead, Smith & Mar (1990)]{1990MNRAS.246..545P} 
Pettini, M., Hunstead, R. W., Smith, L. J., Mar, D. P. 1990, MNRAS, 246, 
545 
\bibitem[\protect\citename{}]{} Pettini M., Steidel C.C., Adelberger
K.L., Dickinson D., Giavalisco M. 2000, ApJ, 528, 96
\bibitem[\protect\citename{}]{} Ponman T.J., Cannon D.B., Navarro
J.F. 1999, Nature, 397, 135
\bibitem[\protect\citename{}]{} Press W.H., Schechter P., 1974, ApJ,
187, 425
\bibitem[\protect\citename{}]{} Quinn T., Katz N., Efstathiou G.,
1996, MNRAS, 279, 49
\bibitem[\protect\citename{}]{} Rauch M., et al., 1997, ApJ, 489, 7
\bibitem[\protect\citename{}]{} Rauch M., 1998, ARA\&A, 36, 267
\bibitem[\protect\citename{}]{} Rees M.J., Ostriker J.P., 1977,
MNRAS, 179, 541
\bibitem[\protect\citename{}]{} Renzini A., 1997, ApJ, 488, 35
\bibitem[] {} Schaye J., , 2000a, in preparation
\bibitem[]{} Schaye J., Rauch M., Sargent W.L.W., Kim, T.-S., 2000a,
ApJL, 541, 1
\bibitem[\protect\citename{}]{} Schaye J., Theuns T., Rauch M., Efstathiou G., Sargent W.L.W.,
2000b, MNRAS, 318, 817
\bibitem[\protect\citename{}]{} Shaver P.A., Wall J.V., Kellermann K.I., Jackson C.A., Hawkins
M.R.S., 1996, Nature, 384, 439
\bibitem[\protect\citename{}]{} Steidel C.C., Giavalisco M., Pettini M., Dickinson M., Adelberger
K.L., 1996, ApJL, 462, 17
\bibitem[\protect\citename{}]{} Stengler-Larrea E.A., Webb J.K., 1993,
in Chincarini G., Ionino A., Maccacaro T., Maccagni D., edts., ASP
Conf. Ser., 51, Observational Cosmology, Astron. Soc. Pac., San
Francisco, p. 591
\bibitem[] {} Tegmark M., Silk J., Evrard A., 1993, ApJ, 417, 54
\bibitem[\protect\citename{}]{} Theuns T., Leonard A., Efstathiou G.,
1998a, MNRAS, 297, L49
\bibitem[\protect\citename{}]{} Theuns T., Leonard A., Efstathiou G.,
Pearce F.R., Thomas P.A., 1998b, MNRAS, 301, 478
\bibitem[\protect\citename{}]{} Thoul A.A., Weinberg D.H., 1996, ApJ,
465, 608
\bibitem[\protect\citename{}]{} Wang B., 1995a, ApJL, 444, 17
\bibitem[\protect\citename{}]{} Wang B., 1995b, ApJ, 444, 590
\bibitem[]{} Webb J.K., 1987, PhD thesis, Univ. Cambridge
\bibitem[\protect\citename{}]{} Wechsler R.H., Gross M.A.K.,  Primack J.R., Blumenthal G.R., Dekel
A., 1998, ApJ, 506, 19
\bibitem[\protect\citename{}]{} Weil M. L., Eke V. R., Efstathiou
G., 1998, MNRAS, 300, 773
\bibitem[\protect\citename{}]{} White S.D.M., Rees M.J., 1978, MNRAS, 183, 341
\bibitem[\protect\citename{}]{} Wu K.K.S., Fabian A.C., Nulsen P.E.J., 1999b, preprint (astro-ph/9910122)
\bibitem[\protect\citename{}]{} Wu K.K.S., Fabian A.C., Nulsen P.E.J.,
2000, MNRAS, 318, 889
\bibitem[\protect\citename{}]{} Zhang Y., Anninos P., Norman M.L.,
1995, ApJL, 453, L57

\end{thebibliography}
\end{document}